\documentclass[aps,epsf,amsfonts,eqsecnum,nofootinbib]{revtex4}
\usepackage{graphicx}
\usepackage{epsfig}
\usepackage{amsmath}
\usepackage{amssymb}
\usepackage{amsthm}
\usepackage{graphics}
\newcommand{\beq}{\begin{equation}}
\newcommand{\beqn}{\begin{eqnarray}}
\newcommand{\eeq}{\end{equation}}
\newcommand{\eeqn}{\end{eqnarray}}

\newcommand{\lp}{\left}
\newcommand{\rp}{\right}
\newcommand{\la}{\langle}
\newcommand{\ra}{\rangle}

\newcommand{\s}{k}
\newcommand{\q}{n}
\newcommand{\z}{ z }

\newcommand{\eff}{{\rm eff}}
\newcommand{\mmu}{\nu}

\newcommand{\uu}{{\cal U}}
\newcommand{\tmn}{\la T_{\mu\nu}\ra}

\newcommand{\kk}{\textsc{kk}}
\newcommand{\bs}{{bs}}

\begin{document}

\title{Quantum fluctuations in the DGP model \\
and the size of the cross-over scale}

\author{ Oriol  Pujol{\`a}s\footnote{pujolas@ccpp.nyu.edu}}

\address{Center for Cosmology and Particle Physics\\
Department of Physics, New York University, New York, NY, 10003,
USA}

\begin{abstract}

The Dvali-Gabadadze-Porrati model introduces a parameter, the
\emph{cross-over} scale $r_c$, setting the scale where higher
dimensional effects are important. In order to agree with
observations and to explain the current acceleration of the
Universe, $r_c$ must be of the order of the present Hubble radius.
We discuss a mechanism to generate a large $r_c$, assuming that it
is determined by a dynamical field and  exploiting the quantum
effects of the graviton. For simplicity, we consider a scalar
field $\Psi$ with a kinetic term on the brane instead of the full
metric perturbations.
We compute the Green function and the 1-loop expectation value of
the stress tensor $\tmn$ of $\Psi$ on the background defined by a
flat bulk and an inflating brane (self-accelerated or not). We
also include the flat brane limit. The quantum fluctuations of the
bulk field $\Psi$ provide an effective potential for $r_c$.
For a flat brane, the 1-loop effective potential $V_\eff(r_c)$ is
of the Coleman-Weinberg form, and admits a minimum for large $r_c$
without fine tuning.
When we take into account the brane curvature, a sizeable
contribution at the classical level changes this picture. In this
case, the potential can develop a (minimum) maximum for the (non-)
self-accelerated branch.

\end{abstract}

\maketitle

\section{Introduction}

One of the most startling observations in cosmology in the last
few years is that our universe is currently in a phase of
accelerated expansion \cite{perl,riess}. One possibility to
explain this phenomenon is to modify the gravitational sector at
large distances.
An interesting realization of this idea was proposed by
Dvali-Gabadadze-Porrati (DGP) \cite{dgp}, and arises within the
Brane World scenario.
In the DGP model, there are two curvature terms, one in the bulk
and the other on the brane. The ratio between the respective
Newton constants $1/M^3$ and $1/m_P^2$ defines an important
quantity, the crossover scale $r_c$.
At distances shorter than $r_c$, gravity behaves as in four
dimensions whereas for larger distances it behaves as in  five
dimensions.\\

This model admits a 'self-accelerated' solution, where the brane
accelerates at late times even with vanishing brane tension or
bulk cosmological constant \cite{cedric}. In this solution, the
acceleration arises as a purely gravitational effect, and the
Hubble rate is $1/r_c$.
On the other hand, $r_c$ must be of the order of the present
Hubble radius in order not to conflict observations
\cite{leaking,DeffayetSNCMB,Maartens,bertolami} (see also
\cite{lue} for a recent review on the phenomenology of the DGP
model). Thus, this model provides an appealing explanation of the
current acceleration of the Universe.
There exists an abundant literature concerning the stability of
the self-accelerated solution \cite{nr,lpr,koyama,gks,cgkp,dRt}.
So far, there seems to be agreement that perturbatively, the DGP
model has ghosts for the self accelerating solution. However, this
mode is strongly coupled, so it is not clear if it remains a ghost
at the full non-perturbative level.
Here, we shall leave this issue aside and, instead, we discuss
another essential ingredient for the model to account for the
current accelerated expansion, namely, whether it is possible to
have a  large $r_c$ naturally.\\

One argument to obtain a large $r_c$ is that the quantum effects
from the numerous matter fields on the brane would generate the
localized kinetic term for the graviton with a large coefficient
\cite{dgp,power}. To this end, though, one would need an enormous
number of fields. Another argument is that the cutoff on the brane
and in the bulk could actually be different \cite{scales}. We
shall also mention that in the model of \cite{gs}, the
corresponding cross-over scale is naturally large. Here, we shall
take the approach that $r_c$ is determined by some (scalar) field,
whose dynamics drives $r_c$ to a large value. Some mechanisms of
this sort were already introduced in \cite{power}. Our aim is to
elaborate on this idea. We shall concentrate on the possibility
that the quantum effects of the bulk fields (the graviton) provide
such a mechanism, given that this requires a minimal extension of
the model.
The main assumption is that the Newton constant in bulk (or on the
brane) is promoted to a Brans-Dicke-type field $\Phi_{}$, which
determines the value of the crossover scale $r_c$. The idea is
then that the quantum effects from the graviton induce an
effective potential for $\Phi_{}$ that drives $r_c$ to large
values.
As we shall see, the 1-loop effective potential induced by the
bulk graviton is of the form
$$
V_\eff^{1-loop}(r_c)={\alpha-\beta\log( \Lambda r_c)\over r_c^4}
$$
where $\beta$ is a constant of order $10^{-3}$ (see below for
details), $\Lambda$ the cutoff of the theory and $\alpha$ is an
order-one constant. This potential has a minimum at $r_c\simeq
e^{\alpha/\beta}/\Lambda$, which can be naturally very large. This
is a very appealing way to generate a large $r_c$, but is not the
end of story.
The reason is that in the most interesting case when the brane is
accelerating, there is a large contribution at the classical level
due to the curvature terms in the bulk and on the brane. In the
simplest situation, the minimum of the potential occurs at a value
of
$r_c$ too small for phenomenological purposes.\\

As mentioned before, here we discuss the quantum fluctuations in
the DGP model. Specifically, the aim is to estimate the vacuum
fluctuations of the graviton on the background defined by a 4D
inflating brane embedded in a flat 5D space. We shall consider the
two possible branches described in \cite{cedric}, and also the
case when the brane is flat. To simplify the analysis, instead of
the full graviton we shall consider a 5D massless minimally
coupled scalar field $\Psi$. This replacement is motivated because
in an appropriate gauge, the spin-2 metric fluctuations obey the
same equation as a massless field, so the analysis of the Green
function do not differ much. In practice, then, we are ignoring
the brane bending mode of the true DGP model.
The discussion of this sector is quite more complicated because it
is strongly coupled \cite{lpr,nr}, aside from the fact that it is
a ghost in the self-accelerated branch \cite{lpr,nr,koyama,gks}.
On the other hand, quantities like the vacuum expectation value of
the stress tensor $\tmn$ may certainly differ for the true graviton
and the scalar considered here, but still we shall
restrict to the latter for illustrative purposes. \\

On the more technical side, we shall mention that the above result
for the potential seems badly divergent in the $r_c\to0$ limit. As
we show in the body of the paper, this is only an artifact of the
renormalization procedure used. Schemes based on analytic
continuation (dimensional regularization, $\zeta-$function) lead
to the expression above. Using a more physical scheme, for
instance introducing a cutoff, one unveils divergences of the form
$\Lambda^k / r_c^{4-k}$ with  $k=1,2,3,4$,  which are implicitly
'renormalized away' by the schemes based on analytic continuation.
Including these terms, the $r_c\to0$ limit is regular as expected.
Still, for large $r_c$, it makes sense to use the usual rules of
renormalization in terms of the effective field $\varphi=1/r_c$.
As we shall see, this is already suggested by the fact that the
classical ('boundary') effective action \cite{lpr} for
the DGP model is analytic in $1/r_c$ (for large $r_c$).\\

The quantum effects from bulk fields with brane-localized kinetic
terms have been considered previously in a slightly different
context. The authors of \cite{pp} computed the Casimir energy
present in a system of two flat branes, and showed that when the
weight of the localized kinetic terms are different, the Casimir
force could stabilize the distance between the branes $L$. The use
of the Casimir effect to stabilize the size of the extra dimension
was first discussed in \cite{ac,candw}, and in the brane world
setup in \cite{gpt,ft,gr}. The work of  \cite{pp}, is relevant
here because in the one brane limit (when one of the brane is sent
to infinity), their result must be compatible with the effective
potential found here in the limit when the brane is flat.
The $L\to\infty$ limit of the result quoted in \cite{pp} vanishes,
apparently suggesting that the potential $V_\eff(r_c)$ is trivial,
which is of course not the case. The reason for this apparent
discrepancy is that reference \cite{pp} deals with the
$L-$dependent part of the potential (the Casimir energy).\\

This paper is organized as follows. In Section \ref{sec:model} we
introduce the scalar version of DGP model. In Section
\ref{sec:fluct} we work out the fluctuations of the model in the
background defined by the de Sitter brane (in both branches). We
compute the Green function in Sec \ref{sec:green} and $\tmn$ in
Sec \ref{sec:tmn}. In Sec \ref{sec:flat} we discuss the limit when
the brane is flat, and Sec \ref{sec:tmnotb} deals with the stress
tensor on the brane. The reader not interested in the technical
details might skip these sections, and go to Section
\ref{sec:veff} where the effective potential for $r_c$ is derived,
and Sec \ref{sec:largerc}, where the mechanism to obtain a large
$r_c$ is discussed.

\section{The DGP model}
\label{sec:model}

The action for the DGP model is
\begin{equation}\label{dgp}
    S= \int d^{5}x \,\sqrt{g}\; M^3R_{(5)}
    +\int d^{4}x \,\sqrt{h} \lp( m_P^2  R_{(4)} -\tau \rp)~,
\end{equation}
where $M$ and $m_P$ are the 5 and 4-dimensional Planck masses,
$R_{(5)}$ is the bulk Ricci scalar, $g$ and $h$ denote the
determinants of the induced metrics in the bulk and on the brane,
$R_{(4)}$ is the Ricci scalar of the brane geometry, and $\tau$ is
the brane tension.
This model admits two solutions with a line element of the form
\cite{cedric}
\begin{equation}
\label{metric}%
ds^2%
=dy^2+a^2(y)ds^2_{(4)}
\end{equation}
where $ds^2_{(4)}$ is the metric on a de Sitter space of unit
curvature radius and with
\begin{equation}
\label{a}%
a(y)=  H^{-1} \pm |y| 
\end{equation}
where $H$ is the Hubble constant on the brane (located at $y=0$). 
Throughout this paper, the upper sign corresponds to the self
accelerated branch (or simply the '$+$' branch) \cite{cedric}. The
$-$ or 'normal' branch can be visualized as the interior of the
space bounded by a hyperboloid (the brane), whereas the $+$ branch
corresponds to the exterior.
For each branch, the Israel junction conditions require that
\beq\label{junction} %
\pm H = r_c H^2- {\tau\over12M^3}~,
\eeq %
where we introduced the cross-over scale
$$
r_c\equiv{m_P^2\over2M^3}~.
$$
Eq. (\ref{junction}) fixes $H$ in terms of $r_c$ and $\tau$. In
the literature, the term 'self-accelerated' is usually reserved
for the solution with $\tau=0$ (for which $H r_c=1$). Here, we
shall consider all values of $H r_c$, which can be obtained by
appropriately choosing $\tau$ and we call 'self accelerated' any
solution in the $+$ branch.

\subsection{A dynamical cross-over scale}
\label{sec:classpot}

We now extend the original induced gravity term to the most
general Brans-Dicke-type modification of the DGP model,
\beq\label{modifiedDGP} %
    S= \int d^{5}x \,\sqrt{g}\; \lp( \Phi_5 \, R_{(5)} +{\rm kin}\rp)%
+ \int d^4x \sqrt{h}\; \lp( \Phi_4 \, R_{(4)} +{\rm kin}\rp)
\eeq %
where $\Phi_{5,4}$ are \emph{unrelated} bulk and brane fields and
we have omitted their kinetic terms, as we are interested in the
configurations with constant fields only. The important point is
that each BD fields control the value of the Newton constant on
the brane or in the bulk so that $\la\Phi_{5}\ra=M^3$, and
$\la\Phi_{4}\ra=m_P^2$. In terms of them the effective cross-over
scale is give by
\beq\label{rc} %
r_c={\Phi_{4}\over2\Phi_{5}}~.
\eeq %
The equations of motion in the model (\ref{modifiedDGP}) become
\beq\label{junction2} %
V_\eff \equiv %
- 12 \Phi_4 H^2 \pm 12 \Phi_5 H + \tau +\delta
V(\Phi_5,\Phi_4) =0 ~, %
\eeq %
together with $V_\eff'=0$, where a prime denotes here
differentiation with respect to $\Phi_5$ or $\Phi_4$. In Eq.
(\ref{junction2}), $\delta V$ represents the potential responsible
for fixing them (which might include bulk and brane
contributions). Note that the condition (\ref{junction2}) is
nothing but the extension of the Friedmann equation
(\ref{junction}).
The first two terms in (\ref{junction2}) are classical
contributions due the brane and the bulk curvature terms in the
action. The former induces a negative runaway potential for
$\Phi_4$, while the latter is positive for the self-accelerated
branch.
As we shall see, for large $r_c$, the 1-loop effective potential
is analytic in $1/r_c\propto 1/\Phi_4$. Thus, it seems difficult
that the quantum effects can induce a minimum at a large $\Phi_4$.
From now on we shall consider that $\Phi_4=m_P^2$ is fixed, and
instead we try to 'stabilize' $\Phi_5$ using the quantum effect
from the graviton.
Note also that this choice automatically leads us to the four
dimensional Einstein frame, which is the physical one. See
\cite{wb-l} for a recent discussion of the cosmology of a DGP
scenario with a dynamical $\Phi_4$ (and constant $\Phi_5$).

\subsection{Scalar toy model}
\label{sec:toy}

We are interested in the fluctuations of the graviton propagating
in the background (\ref{metric}). One can show that the spin-2
polarizations of the graviton perturbations obey the same equation
of motion as a massless minimally coupled scalar field (see {\em
e.g.} \cite{koyama}) with a kinetic term on the brane. Hence, we
shall simplify the problem and rather consider a scalar field
(even under $Z_2$ symmetry) propagating on the background
(\ref{metric}). Its action is
\begin{equation}\label{action}
    S=-{1\over2}\int d^5x \,\sqrt{g}\;
    (\partial_\mu \Psi)^2
    -\int d^4x \,\sqrt{h} \; r_c \lp(\partial_i\Psi\rp)^2 ~,
\end{equation}
where, $g$ and $h$ denote the determinants of the background
metrics on the bulk and on the brane (\ref{metric})%

\subsection{Classical boundary effective action}
\label{sec:boundact}

Let us now discuss how the cross-over scale appears in the
effective four dimensional theory. Consider for simplicity the
solution with a flat brane, and we have in mind a situation with
large $r_c$ (compared to any other scale in the problem).
A simple way to obtain the effective theory of the model
(\ref{action}) is to compute the effective boundary action.
Following \cite{lpr}, once some data on the brane are specified
for the bulk field $\Psi|_0=\psi$, the equation of motion in the
bulk is solved by $\Psi=e^{-y\Delta}\psi$, where
$\Delta=(-\Box_{(4)})^{1/2}$ and $\Box_{(4)}$ is the 4D
D'Alembertian operator. Inserting this solution back into the bulk
action, we can integrate it explicitly and obtain a surface term.
Thus, the \emph{boundary effective  action} is
$$
S_{\eff}= \int d^4x \sqrt{h} \; \psi\lp( r_c \Box_{(4)} -
 { \Delta } \rp) \psi,
$$
For large values of $r_c$, the four dimensional term dominates.
Hence, it is preferable to express this as the action for a 4D
canonical field. Restricting our attention to constant
configurations of $r_c$, we can always rescale the field as
$\hat\psi=\sqrt{2r_c} \,\psi$ in the canonical form, so the
effective action is
\beq\label{boundaction} %
S_{\eff}= {1\over2}\int d^4x \sqrt{h} \lp( \hat\psi
\Box_{(4)}\hat\psi  - \varphi \,\hat\psi \Delta \hat\psi \rp),
\eeq %
where we introduced
$$
\varphi\equiv {1\over r_c}~.
$$
This effective action describes the resonant or metastable mode of
the DGP model. This mode propagates as in four dimensions during a
typical distance $r_c$ after which it behaves as in five
dimensions. Put another way, the decay width into five dimensional
modes is is of order $1/r_c$. Hence, at distances shorter than
$r_c$ this 4D effective representation is accurate.

The point that we want to emphasize here is that in this regime,
the effective action (\ref{boundaction}) is analytic with respect
to $\varphi$.
This is a rather special feature of this model. As we shall see,
all the explicit expressions computed below are regular in the
$\varphi\to0$ limit.
%
Finally, let us note that in terms of the model of Sec
\ref{sec:classpot} we have $\varphi={2\Phi_5 /m_P^2}$, so we can
interpret $\varphi$ as the bulk Brans-Dicke field (restricted on
the brane).\\

Before switching to the explicit computation of the Green
function, we shall briefly discuss the generalization of the
boundary effective action to the curved brane case.
It is straightforward to show from Eqs. (\ref{eom}) and
(\ref{schrod}) that when the brane is inflating, the solution for
$\Psi$ in the bulk with data on the brane specified by $\psi$ is
given by $\Psi=a^{-3/2}e^{-|z| \hat p}\psi$, where $z$ is the
conformal coordinate ($dz=dy/a$), $\hat p \equiv
(-\Box_{(4)}+(3/2)^2)^{1/2}$ and now $\Box_{(4)}$ is the covariant
D'Alembertian in de Sitter space. Using the same logic as above,
the boundary effective action in this case is
\beq\label{boundactiondS}%
S_\eff= \int d^4x\sqrt{h}\;\psi \lp( r_c
\Box_{(4)}\mp{3\over2}H-\sqrt{\lp({3H\over2}\rp)^2-\Box_{(4)}}
\rp) \psi~.
\eeq%
From the analytic structure of this action as a function of
$\Box_{(4)}$, we can already identify the main features of the
effective theory. For both branches, there is a branch cut
starting at $\Box_{(4)}=(3H/2)^2$, which corresponds to the
continuum of Kaluza-Klein modes. Aside from this, we can also see
the presence of one lighter mode. Performing the same rescaling of
$\psi$ as above, we can identify the mass of this mode as the
non-derivative term in (\ref{boundactiondS}). For the
non-self-accelerated branch, this mode is massless. As we will see
in short, this agrees perfectly with the Kaluza-Klein
decomposition.
For the self-accelerated branch, we read from
(\ref{boundactiondS}) a mass-squared given by $3H/r_c$, which
agrees with the analysis of the KK spectrum (\ref{mbs}) to leading
order (in the large $Hr_c$ limit).

\section{Quantum fluctuations}
\label{sec:fluct}

In this Section, we shall take the bulk spacetime to be $(n+2)$
dimensional. The Klein-Gordon equation on the space
(\ref{metric}),
\beq\label{eom} %
\lp [ \Box_{(n+2)} + Hr_c \delta(z) \Box_{(n+1)} \rp] \Psi =0
\eeq %
is separable into modes of the form
\beq\label{kkdecomp}%
\Psi(z,x^\mu)=\sum_p \uu_p(z)\psi_p(x^\mu)~,%
\eeq%
In the previous equation, $\Box_{(n+2)}$ and $\Box_{(n+1)}$ are
the bulk and brane D'Alembertians. For clarity, we omit the set of
indices labelling the angular momentum of the four dimensional
modes $\psi_p(x^\mu)$.
The $(n+1)-$dimensional modes obey %
$\lp[ \Box_{(n+1)} - m^2 \rp] \psi_p=0$ %
and represent four dimensional fields with mass
$$
m^2\equiv[(n/2)^2+p^2]H^2~.
$$
The mass spectrum is determined by the radial equation, which can
be cast in the Schr\"odinger form
\beq%
\label{schrod}%
\left[-\partial_z^2+V(z)\right]\widetilde\uu_p=p^2\widetilde\uu_p%
\eeq%
with $\widetilde\uu_p=a^{n/2}\uu_p$, and
\beq%
V(z)={\lp(a^{n/2}\rp)''\over a^{n/2}}-{n^2\over 4} - 2  {r_c m^2\over H} \, \delta(z) 
=\lp( \pm n H - 2  {r_c m^2\over H} \, \rp) \,\delta(z)~,
\eeq%
where a prime denote differentiation with respect to $z$, and we
introduced the conformal coordinate $dz=dy/a$. Hence, the modes
are combinations of $e^{\pm i p |z|}$, and the spectrum is
composed of a continuum of Kaluza-Klein modes with masses
$m_\kk>nH/2$, plus possibly one discrete (or
'bound') state with mass $m_\bs<nH/2$.\\

In terms of the mode decomposition,  the boundary condition on the
brane 
reduce to
\beq%
\label{bc}%
\lp[\partial_z-\nu_\pm(p) \rp] \widetilde\uu_p \; \big|_{0+}=0
\eeq%
where $|_{0+}$ stands for $\lim_{z\to0}$ with $z>0$, and
\beq%
\label{nu}%
\nu_\pm(p)\equiv \pm{n\over2}-{\lp( p^2+\lp(n\over2\rp)^2 \rp) H
r_c } ~.
\eeq%
Note that (\ref{bc}) is similar to the Dirichlet boundary
condition for high KK mass modes, whereas it recovers the usual
Neumann-type condition for light modes. This is indeed the key
ingredient to obtain a four dimensional behaviour at short
distances in the DGP model, since it implies that high energy
modes are suppressed on the brane.

We shall normalize the modes according to
\beq\label{norm} %
\lp( \Psi_1,\Psi_{2}\rp)\equiv   \int d\Sigma\;\xi^\mu\ %
i\lp( \Psi^*_1 \partial_\mu \Psi_2 %
-  \Psi_2 \partial_\mu \Psi^*_1  \rp)%
+ 2\  r_c \int d\sigma\;\xi^\mu \, %
i \lp( \Psi^*_1 \partial_\mu \Psi_2 %
-  \Psi_2 \partial_\mu \Psi^*_1  \rp)|_0 
\eeq %
where $\xi^\mu$ is a future-directed timelike unit vector and
$d\Sigma$ ($d\sigma$) is the volume element of the surface
$\Sigma$ ($\sigma$) normal to $\xi^\mu$ in the bulk (on the
brane). For any pair of modes satisfying the equation of motion
(\ref{eom}), the norm (\ref{norm}) is independent of the choice of
$\Sigma$.
We shall take $\xi^\mu$ to be along the time coordinate of the
observers on the brane. It is convenient to introduce the Rindler
coorinates $\{a,t\}$, related to the
rectangular coordinates in the bulk $\{T,\bf{X}\}$ through %
\begin{align}%
 R&= a \cosh t\cr
 T&= a \sinh t
\end{align}%
where $R=|\bf{X}|$.
In terms of these coordinates,  $\xi^\mu = \delta^\mu_t/a$ and %
the metric on the de Sitter slices $y=const$ %
are $ds^2_{(n+1)}= a^2[-dt^2+ \cosh^2 t \,d\Omega^2_n]$ where
$d\Omega^2_{(n)}$ is the metric on an $n-$dimensional sphere.

For a mode of the form $\Psi_{p,l}=\uu_{p}\psi_{p,l}$ where $l$ is
a collective index of the dS part, we have
\beqn\label{norm2} %
\lp( \Psi_{p,l}\,,\Psi_{p,l} \rp)%
&=& \int dz
\,a^n(z) \lp( 1+2{r_cH} \delta(z) \rp)\;%
|\uu_p|^2  \lp( \psi_{p,l},\psi_{p,l}\rp)_{(n+1)} ~,
\eeqn %
where $\lp(\psi_1 ,\psi_2 \rp)_{(n+1)}$ is the usual
four-dimensional Klein Gordon product, %
$\int d\sigma\;\xi^\mu\,i\lp( \psi^*_1 \partial_\mu \psi_2 %
-  \psi_2 \partial_\mu \psi^*_1  \rp)$.
The continuum of KK modes is normalized as
$(\Psi^\kk_{p,l},\Psi^\kk_{p',l'})=\delta(p-p')\delta_{l,l'}$,
which leads to
$$
\int_{-\infty}^{\infty} dz (1+2Hr_c
\delta(z))|\uu_\kk|^2=\delta(p-p')~,
$$
while for the bound state we shall impose
$$
\int_{-\infty}^{\infty} dz (1+2Hr_c \delta(z))|\uu_\bs|^2=1~.
$$

A bound (or 'localized') state is a mode with pure imaginary
$p=i\lambda$ and $\lambda>0$, so that its wave-function
$\widetilde \uu_\bs\sim\,e^{-\lambda |z|}$ is normalizable. The
boundary condition implies that $\lambda$ must satisfy
\beq\label{bc2} %
-\lambda-\nu_\pm(i\lambda)=0
\eeq %
where $\nu_\pm$ is given in (\ref{nu}). The two roots are, for
each branch,
\beqn %
\lambda_1 &=&  \mp { n \over 2 } \cr %
\lambda_2 &=& \pm {n\over2}-{1\over H r_c} %
\eeqn %
It is convenient to introduce the notation $\lambda_>$ and
$\lambda_<$ as the larger and smaller of $\lambda_{1,2}$
\footnote{To be precise, here we are assuming that $Hr_c$ is large
enough}.
%
%
Only $\lambda_>$ is positive,
so this root corresponds to the bound state in both
branches. Its wave-function is $\widetilde
\uu_\bs=N_\bs\,e^{-\lambda_> z}$
with (see (\ref{norm}))
\beq\label{n2bs} %
N_\bs^2={\lambda_>\over 1+2Hr_c\,\lambda_>}={\lambda_>\over n
Hr_c\mp1} ~.
\eeq %
In the $-$ branch, the bound state is massless ($\lambda_>=n/2$).
In the self accelerated branch, it is normalizable only for $Hr_c> 2/n$,
and its mass is 
\beq\label{mbs} %
m^2_{\bs(+)}={n H r_c-1\over r_c^2}~.
\eeq %
The behaviour of this mode as a function of $r_c$ is as follows.
For $Hr_c\gg 2/n$, the bound state is light. For smaller $r_c$, it
acquires a mass until at $r_c=2/nH$ it degenerates with the KK. At
this point, this becomes unnormalizable and it should be rather
described as a resonance, or quasi-normal mode. Its mass acquires
an imaginary part $p=p_0-i\Gamma$ that is interpreted as the decay
width into KK modes \cite{rubakov,ls}. In this model, this
actually turns out to be a purely-decaying quasi-normal mode
($p_0=0$), with decay width $\Gamma=1/ r_c-3H/2$.

The other mode, at $\lambda_<$, is not normalizable, so strictly
speaking it is not included in the spectrum. However, one can
think that it corresponds to a resonance, or equivalently that the
whole KK continuum in some sense behaves like this mode. For the
$-$ branch, its decay width is $1/r_c+3H/2$, and for the $+$
branch it is $3H/2$. Note that in the latter case, we can have two
different resonances for $Hr_c< 2/n$. The spectra of normalizable
and resonant modes is sketched in Fig. (1).

\begin{figure}[tb]
$$
\begin{array}{cc}
  \includegraphics[width=9cm]{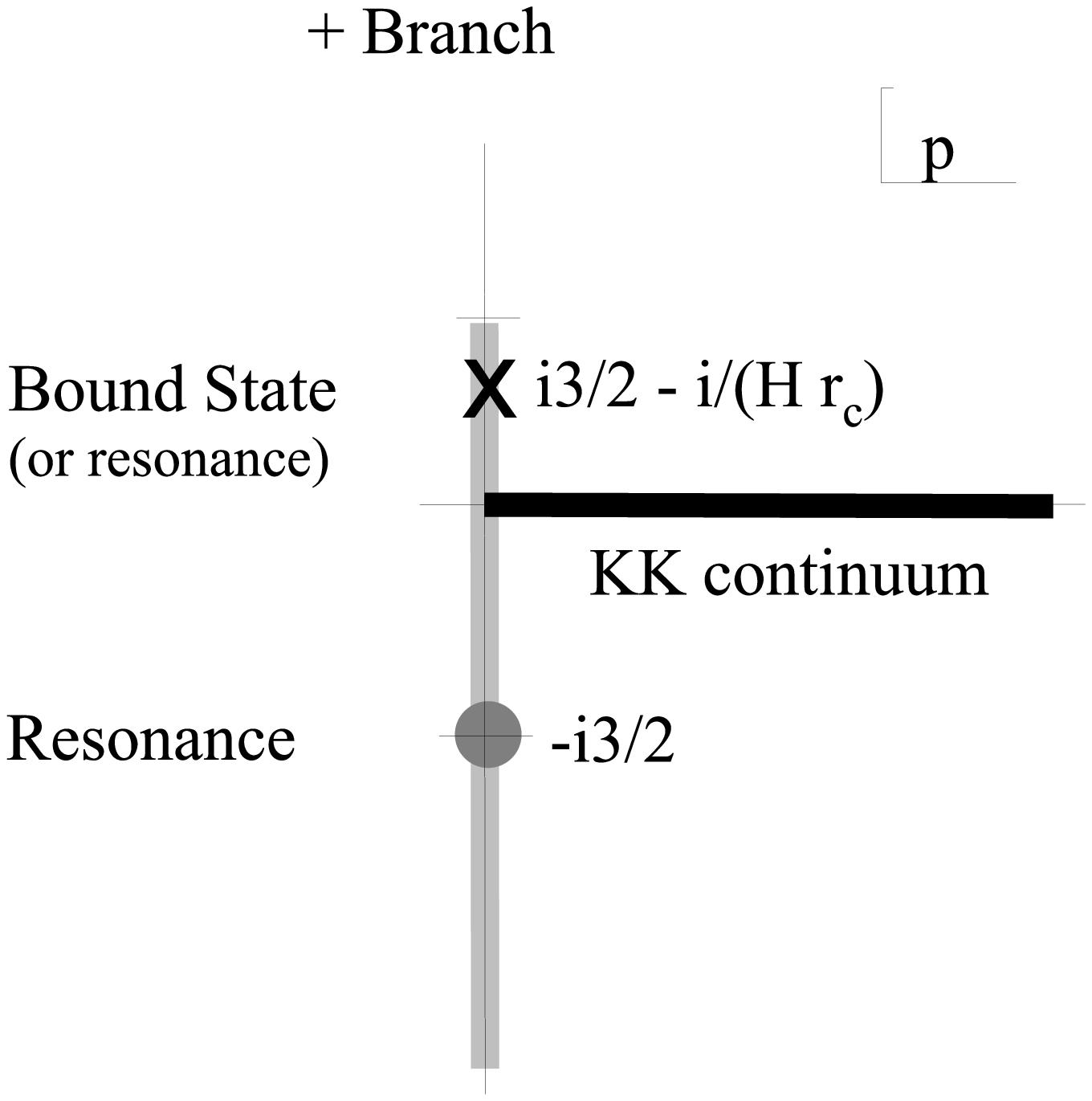}
  &\includegraphics[width=9cm]{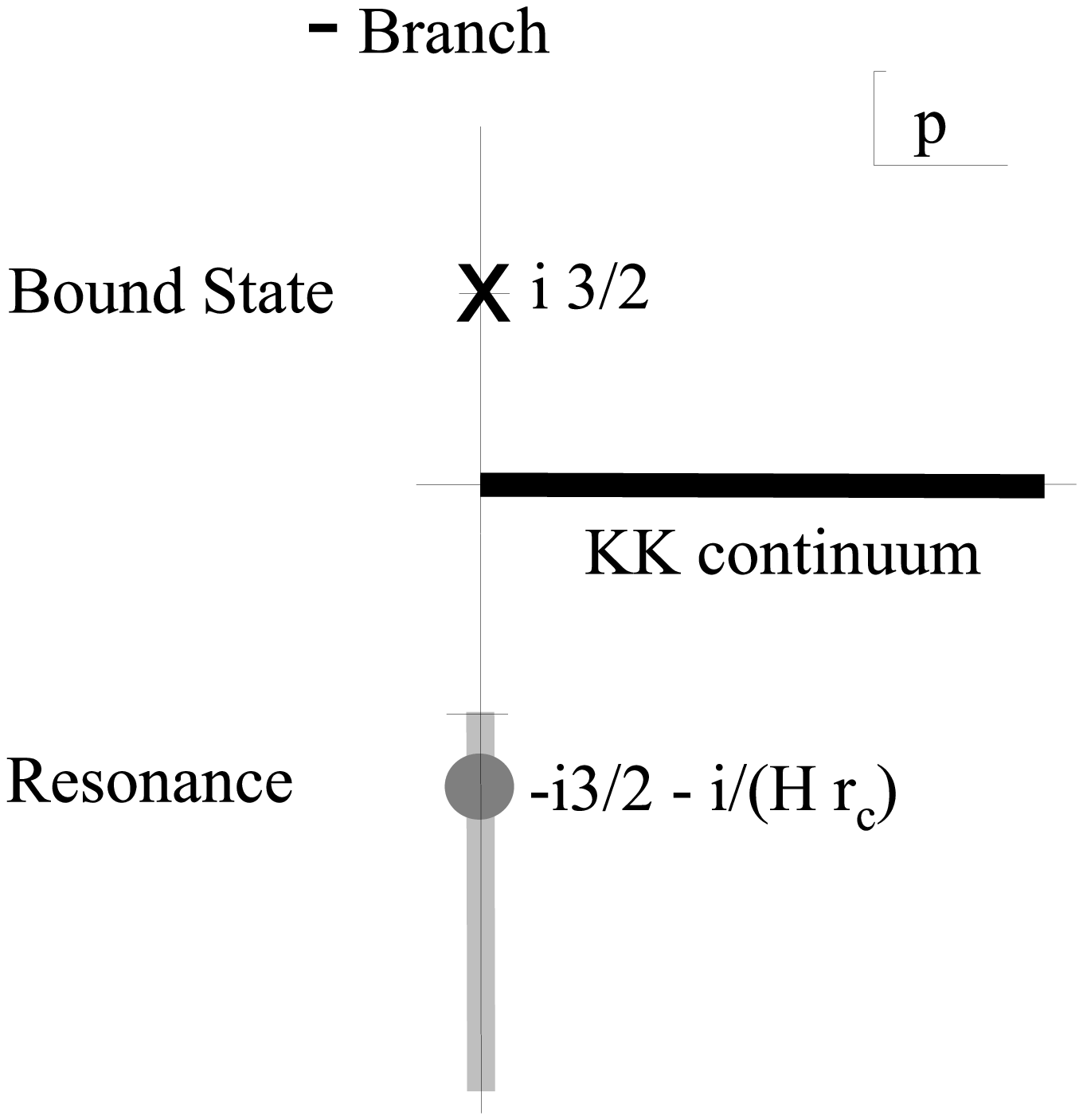}\nonumber\\[-.5cm]
  {\rm (a)}& {\rm (b)}
\end{array}
$$
\label{fig:poles} %
\caption{We represent the spectrum of modes in the $p$ plane,
related to the mass by $m^2=(n^2/4+p^2)H^2$. The continuum of KK
modes have $p$ real and are represented by the thick black line,
both for the self-accelerated branch, (a), and for the
non-self-accelerated branch, (b). The crosses denote the positions
of the other two discrete modes in the spectrum for some
representative value of $Hr_c$. In the upper half plane, these
modes are normalizable and correspond to the bound state. In the
lower half plane, they are not normalizable, and represent a
resonance. The shaded line indicates the possible values that the
pole at $p=i\lambda_2$ can take as $Hr_c$ is varied from $0$ to
$\infty$. Note that in the self accelerated branch and for
$Hr_c<2/n$ there is no bound state. Rather, there are two
resonances.}
\end{figure}

\subsection{The Green function}
\label{sec:green}%

Before starting the computation of the Green function on the
backgrounds (\ref{metric}), we shall mention that for a massless
minimally coupled field, there is an ambiguity in the definition
of the Green function in an overall additive constant. The shift
symmetry $\Psi\to\Psi+const$ allows us to 'gauge away' any
constant contribution. This turns out to be important for the
non-self accelerated branch because in this case there is a
massless 4D mode, and as is well known massless minimally coupled
scalar fields enjoy an (IR) divergent Green function in spaces
with closed spacelike sections. In any case, and precisely thanks
to the shift symmetry, this divergence does not appear in the
stress tensor $\tmn$.
This issue has been discussed at length in four dimensions
\cite{af,gk} and in the brane-world context \cite{pt}.
These explicit computations further show that both $\tmn$ and the
Green function, are everywhere regular.

We can compute the (Hadamard) Green function as a mode sum,
\begin{equation}\label{greg}
\begin{array}{rcll}
  G_{}^{(1)}&=&G^{\kk}+G^{bs}~, & \textrm{with} \\[2mm]
  G^{bs}&=&\theta(\lambda_>)\,\uu^\bs(z)\uu^\bs(z')\,G_{i\lambda_>\,(dS)}^{(1)}&\textrm{and} \\[2mm]
  G^{\kk}&=&\displaystyle\int_0^\infty dp \;\uu^\kk_p(z )
      \uu^\kk_p(z')\;G_{p\,(dS)}^{(1)}~,&
\end{array}
\end{equation}
where $G_{p\,(dS)}^{(1)}$ denotes the massive Bunch-Davies Green
function in $(n+1)-$ dimensional de Sitter space. The
$\theta(\lambda_>)$ factor in $G^\bs$ ensures that it is only
included when it is normalizable.

One easily finds \cite{pt} (for the 'right' half of the space,
$z>0$) that the normalized radial KK modes are
\begin{equation}\label{uukk}
 \uu^\kk_p(z)=\sqrt{1\over \pi a^{\q}(z)
   (1+(\mmu/p)^2)}\left(
   \cos \left( p\z  \right)
         +{\mmu\over p}\sin \left( p\z   \right)\right)~,
\end{equation}

It is convenient to introduce the \emph{subtracted} Green function
\beq\label{gsub1} %
\overline{G}^{(1)}\equiv G^{(1)}-G^{(1)}_{Mink}~,
\eeq %
where
$G^{(1)}_{Mink}$ is the Green function in Minkowski space, as it
is free from ultraviolate divergences in the bulk. In terms of the
mode expansion,
\beq%
\label{gsub}%
\overline{G}^{(1)}= G^\bs+\int_{-\infty}^\infty dp
\lp[\uu^\kk_p(z)
\uu^\kk_p(z')\rp]^{sub}G_{p\,(dS)}^{(1)} %
\eeq%
with
\begin{equation}
\label{uu}
\lp[\uu^\kk_p(z) \uu^\kk_p(z')\rp]^{sub}%
   ={1\over 4\pi \lp(a(z)a(z')\rp)^{n/2} }
    { p -i\nu\over p+i\mmu}e^{i(z+z')p}~.
\end{equation}
One can now perform the $p$ integral in (\ref{gsub}) closing the
contour in the complex $p$ plane by the upper half plane and
summing the residues of the enclosed poles.
Using
\beqn %
\label{fractions}
{p-i\mmu\over p+i\mmu}
&=&-1-2 i{\lambda_>+\lambda_<\over \lambda_>-\lambda_<}\lp(
{\lambda_>\over p-i\lambda_>} -{\lambda_<\over p-i\lambda_<}\rp)
\eeqn%
%
%
we see that the only pole from $\lp[\uu^\kk_p(z)
\uu^\kk_p(z')\rp]^{sub}$ contributing to (\ref{gsub}) is at
$p=i\lambda_>$, the bound state location. Given that
$$
{\lambda_>+\lambda_<\over \lambda_>-\lambda_<}=-{1\over nHr_c
\mp1}~,
$$
we find that this pole in the contribution from the KK modes
equals the contribution from the bound state $G^\bs$ (see Eq.
(\ref{n2bs})), with the opposite sign \footnote{Note that any
normalization of the modes other than (\ref{norm}) would not lead
to this cancellation, with the corresponding wrong short distance
behaviour of the Green function.}.
%
%
%
%
The remaining poles in (\ref{gsub}) arise from
$G_{p\,(dS)}^{(1)}$. Performing the sum over them, (see the
Appendix for details) we find
\beqn %
\label{convenient}%
\overline{G}^{(1)}
&=&-G_{0}
-2{\lambda_>+\lambda_<\over \lambda_>-\lambda_<}\lp(\lambda_> \;%
G_{\lambda_>}
-\lambda_<\;%
G_{\lambda_<}
\rp)
\eeqn %
where
\beqn %
\label{G0}
  G_0 &\equiv&%
{2H^n\over n S_{(n+1)}}   \left({1\over 1 + (aa'H^2)^2 - 2 aa'H^2
\cos\zeta} \right)^{\q /2}~,
\eeqn %
\beqn \label{op2} %
G_\lambda&\equiv&{2H^n\over n S_{(n+1)}} %
{\lp(aa'H^2\rp)^{-n{(1\pm1)\over2}} \over n/2-\lambda}\; %
\;F_1\lp( {n\over2}-\lambda,{n\over2},{n\over2},{n\over2}-\lambda+1; %
\; \lp(aa'H^2\rp)^{\mp1} e^{i\zeta}\;, \lp(aa'H^2\rp)^{\mp1} e^{-i\zeta}\rp)%
\eeqn%
Here, $\zeta$ is the invariant distance in $dS$ space and
$S_{(n+1)}=2 \pi^{1+n/2}/\Gamma(1+n/2)$ is the volume of a unit
$n+1$ dimensional sphere, and
\beq\label{appell} %
F_1\lp( a,b_1,b_2,c; z_1,z_2 \rp)=\sum_{k=0}^\infty \sum_{l=0}^\infty%
{(a)_{k+l} (b_1)_{l} (b_2)_{k}\over (c)_{k+l}}
{z_1^l\, z_2^k\over l!k!}
\eeq %
is the Appell Hypergeometric function of two variables (here,
$(a)_k\equiv \Gamma(a+k)/\Gamma(a)$).

In the $-$ branch $\lambda_>=n/2$, so it is clear from (\ref{op2})
and (\ref{appell}) that the $\lambda_>$ term in (\ref{convenient})
induces a divergence, as previously anticipated. This divergence
arises from the constant mode of the bound state, which is
massless in this branch. It is well known that massless scalar
fields in de Sitter space produce this type of infrared
divergence. See \cite{af,gk} for a discussion in dS space, and
\cite{pt} for a similar brane model. Here, we will simply ignore
all constant contributions to the Green function, given that they
do not contribute to any observable associated with $\Psi$. This
is a consequence of the shift symmetry of the model,
$\Psi\to\Psi+const$, which in other words allows us to gauge away
the constant contributions. Hence, the leading order behaviour of
(\ref{convenient}) near the light cone ($aa'\to0$) is
$$
\overline{G}^{(1)}(y,y',x,x')
\sim   H^{n+2} aa' \cos\zeta ~.
$$
For a source point $x'$ on the brane with $t=0$, and $x$ close to
the horizon ($U\equiv R-T=0$), then $\zeta$ is large and pure
imaginary, so we have $\overline{G}^{(1)} \sim   H^{n+1} T$ ,
which is perfectly regular on the horizon.

For the $+$ branch, we find that at infinity ($aa'\to\infty$) the
leading order terms in (\ref{convenient}) cancel each other, and
we have
$$
\overline{G}^{(1)}(y,y',x,x)
\sim H^n\,(aa' H^2 )^{-(n+1)} \cos\zeta~.
$$
The disturbance generated by a source on the brane at $t'=0$, at
spatial infinity decays slightly faster than for a flat brane,
$\overline{G}^{(1)} \sim 1/(H y^{(n+1)}) $. See the Appendix for
the form of $\overline{G}^{(1)}$ in the bulk in the coincidence
limit.

\subsection{Vacuum expectation value of the stress tensor}
\label{sec:tmn}

The stress tensor for a classical field configuration can also be
split into bulk and a surface parts as
\begin{eqnarray}\label{tmnclass}
T_{\mu\nu}^{bulk}&=&
\partial_\mu\Psi\partial_\nu\Psi-{1\over2}\;
\left(\partial_\rho\Psi\right)^2 g_{\mu\nu} \cr
T_{ij}^{brane}&=&2r_c\,\lp[\partial_i\Psi\partial_j\Psi-{1\over2}\;
\left(\partial_k\Psi\right)^2 h_{ij} \rp]\,\delta(y)~,
\end{eqnarray}
where $h_{ij}$ is the induced metric on the brane, and the
equation of motion has been used.
In point splitting regularization, the v.e.v. $\tmn$  is computed
as \footnote{We omit the anomaly term since it vanishes in the
bulk for odd dimension, and on the brane it can be absorbed in a
renormalization of the brane tension.}
\begin{eqnarray}
\label{splitting}
\la T_{\mu\nu}\ra^{bulk}&=&{1\over2} \lim_{y',x'\to y,x}\; \lp\{
\partial_\mu\partial'_\nu-{1\over2}\; g_{\mu\nu}
g^{\lambda\sigma}\partial_\lambda\partial'_\sigma \rp\} \overline{G}^{(1)}(y,x;y',x')%
\qquad\textrm{and}\nonumber\\[1mm]%
\la T_{ij}\ra^{brane}&=&r_c \,\delta(y)\,%
\lim_{x'\to x}\; \lp\{ \partial_i\partial'_j-{1\over2}\; h_{ij}
h^{kl}\partial_k\partial'_l \rp\} %
\overline{G}^{(1)}(0,x;0,x')~,
\end{eqnarray}
where $\partial'_\mu=\partial/\partial x'^\mu$.

For the $-$ branch, we find
\begin{eqnarray}
\label{Tyy-}%
\la T^y_{~y}\ra^{bulk} =  {n\over2 S_{(\q+1)} } H^{n+2}
    \Biggl\{
        -{1\over (1-(aH)^2)^{\q +1} }
        &-& {1\over 1+nHr_c} {1\over (aH)^2} ~
        {(1-(aH)^2)^n -1 \over (1- (aH)^2)^n}\\
        &-&{n H r_c+2\over Hr_c(1+nHr_c)}\, \lp(aH\rp)^{-2(n+1+1/Hr_c)} \;%
        B_{(aH)^2}\left(n +1+{1\over Hr_c},-n\right)%
        ~\Biggr\}\nonumber
\end{eqnarray}
From the local conservation of the stress tensor, the remaining
components $\la T^i_{~j}\ra \equiv -\rho \,h^i_{~j} $, are given
by
$$
\rho^{bulk}=-\lp(1+{1\over n+1}\,a\partial_a\rp) \la
T^y_{~y}\ra^{bulk}~.
$$
Note that, on the light cone ($a=0$), $\rho\simeq-\la T^y_{~y}\ra$
and approach a constant value of order $H^{n+2}$. Thus, as
expected, $\tmn$ is completely regular on the horizon.
%

For the $+$ branch, we find
\begin{eqnarray}
\label{Tyy+}%
\la T^y_{~y}\ra^{bulk} =  {n\over2 S_{(\q+1)}} H^{\q +2}
    \Biggl\{
        {1\over ((aH)^2-1)^{\q +1} }
        &-& {1\over 1-nHr_c} {1\over (aH)^{2(n+1)}} ~
        {1 \over (1- 1/(aH)^2)^n}\\
        &-&{nHr_c-2\over Hr_c(1-nHr_c)}\, \lp(aH\rp)^{-2(n+1-1/Hr_c)} \;%
        B_{1/(aH)^2}\left({1\over Hr_c},-n\right)%
        ~\Biggr\}\nonumber~.
\end{eqnarray}
Note that in spite of appearances, this expression is finite for
$Hr_c=1/n$.
At infinity ($a\simeq y \to\infty$), the leading order terms of
the three summands in (\ref{Tyy+}) cancel each other, so $\la
T^y_{~y}\ra^{bulk}$ and $\rho^{bulk}$ decay like
$H^{n+2}(aH)^{-2(n+2)}\simeq {1/   (H y^2)^{(n+2)}}$.

Close to the brane ($y\to0$) ,
$$
\la T^y_{~y}\ra^{bulk} \simeq  \pm {n\over2 S_{(\q+1)} }  %
 {H\over y^{n+1}}\lp( 1+ {\cal O} \lp(Hy\rp) \rp) ~,
$$
so $\la T^y_{~y}\ra^{bulk} $ vanishes in the $H\to0$ limit, as
required by the local conservation of the stress tensor.
%
As for the components parallel to the brane  (for $n=3$), we find
\beq\label{Tyyy=0}%
\la T^i_{~j}\ra^{bulk} =  {9\over256\pi^2}\lp({1\over y^5}%
\mp{2+3 Hr_c\mp9 (Hr_c)^2\over2(1\mp3Hr_c) }{1\over r_c y^4}%
+\dots\rp) \;h^i_{~j}~.
\eeq%
The $r_c\to0$ limit of this expression is not well defined because
it holds only for $y$ smaller than $H$ and $r_c$.

\subsection{Flat brane limit} \label{sec:flat}

To recover flat brane case, we take $\zeta\simeq H \Delta x$ where
$\Delta x$ is the distance between points $x$ and $x'$, in $n+1$
dimensional flat space. If we take $H\to0$ and keep $Hr_c$ (and
hence $\lambda_\gtrless$) fixed, this means that we are letting
$r_c\to\infty$, and will obtain the limit with Dirichlet boundary
conditions.
Recalling that in terms of the proper bulk coordinate $y$ we have
$aH=e^{\pm z}=1\pm H y$ and taking the limit $H\to0$ in
(\ref{convenient}), (\ref{G0}) and (\ref{op2}), one easily obtains
$$
\overline{G}^{(1)}\lp(y,y';x,x'\rp)= -{2\over n S_{(n+1)}} {1\over
\lp(Y^2+\Delta x^2\rp)^{n/2}}~.
$$
where $Y\equiv y+y'$, which indeed corresponds to the renormalized
Hadamard function in flat space with Dirichlet boundary
conditions.

To recover the $H\to0$ limit keeping $r_c$ finite, we have to take
into account that in this limit,\footnote{The contribution from
$\lambda_1$ in (\ref{convenient}) vanishes in the limit $H\to0$}
$\lambda_2\to-\infty$.
Using the representation
$$
{1\over
a}F_1\lp(a,b,b,a+1;z_1,z_2\rp)=\int_0^\infty\;d\tau\;%
{e^{-\tau} \over \lp( (1-z_1\,e^{-\tau/a})(1-z_2\,e^{-\tau/a})
\rp)^b} ~,
$$
we obtain
\beqn %
\lim_{\lambda\to-\infty}\;-\lambda\; 
G_{\lambda} %
&=&{2\over n S_{(n+1)}} %
{1\over r_c^n}\int_0^\infty d\tau\; %
{e^{-\tau} \over \lp( \tau^2+ (Y^2+\Delta x^2)/r_c^2-2Y/r_c \rp)^{n/2}} ~.%
\eeqn %
Specializing to $n=3$, we obtain at $Y=0$,
\beqn %
\lim_{H\to0}\; \overline{G}^{(1)}\lp(0,0;x,x'\rp)= {1\over
4\pi^2}\lp( -{1\over \Delta x^{3}}
%
-
{\pi \over  r_c^2} { Y_1\lp(\Delta x/r_c\rp) %
+H_{-1}\lp(\Delta x/r_c \rp)  %
\over \Delta x } %
\rp)~, %
\eeqn %
where $Y_\nu(z)$ is the Bessel function of the second kind and
$H_\nu(z)$ is the Struve function.%
On the other hand, for $Y\neq0$ and $\Delta x=0$, we have
$$
\lim_{H\to0}\; \overline{G}^{(1)}\lp(y,y';x,x\rp)= {2\over n
S_{(n+1)}}\lp( -{1\over Y^{n}} + {2\over r_c^n}
\;e^{Y/r_c}\Gamma\lp(1-n,{Y\over r_c}\rp)\rp)
$$
Note that in the limit $r_c\to0$, we obtain
$\overline{G}^{(1)}\lp(y,y';x,x\rp)=1/(n S_{(n+1)} Y^n)$,
recovering the result for an ordinary scalar with Neuman boundary
conditions. Conversely, in the  $r_c\to\infty$ limit, we obtain
$\overline{G}^{(1)}\lp(y,y';x,x\rp)=-1/(n S_{(n+1)} Y^n)$,
corresponding to Dirichlet boundary conditions.

~\\

As for the vacuum expectation of the stress tensor, one easily
finds that $\la T^y_{~y}\ra^{bulk}=0$ as required by the local
conservation
of the stress tensor, and%
\begin{eqnarray}
\label{TijFlat}%
\la T^i_{~j}\ra^{bulk} &=&  {n\over S_{(\q+1)}}
    \lp[{1\over (2y)^{n+2}}
    -{2\over r_c^{n+2} }\;e^{2y/r_c}\;\Gamma\lp(-1-n,{2y\over
    r_c}\rp)\rp]\;h^i_{~j}~.
\end{eqnarray}
Near the brane, the stress tensor behaves like
\begin{eqnarray}
\label{TyyFlaty=0}%
\la T^i_{~j}\ra^{bulk} &\simeq&  {9\over 256\pi^2}\lp({1\over
y^5}-{1\over r_c y^4}+{2\over 3}{1\over r_c^2 y^3}-{2\over
3}{1\over r_c^3 y^2}+{4\over 3}{1\over r_c^4 y}+{8\over
3}{\log\lp(y/r\rp)\over r_c^5 }+\dots\rp)\;h^i_{~j}~.
\end{eqnarray}
The first term of this expression agrees with \cite{romeosaharian}
and with the $H\to0$ limit of Eq. (\ref{Tyyy=0}). Note that the
agreement beyond the first term holds for the self accelerated
branch only.

On the other hand, at large distances $y\gg r_c$,
\begin{eqnarray}
\label{TyyFlaty=0}%
\la T^i_{~j}\ra^{bulk} &=&  -{9\over 256\pi^2}{1\over
y^5}\;h^i_{~j}\lp(1 +{\cal O}\lp(r/y\rp)\rp)~.
\end{eqnarray}
One observes that it behaves like close to the brane but with the
sign flipped. This happens because the $y\gg r_c$ ($y\ll r_c$)
regime is equivalent to the $r_c\to0$ ($r\to\infty$) limit. In
those limits, the field $\Psi$ is subject to Neuman or Dirichlet
boundary conditions respectively. As is well known, the sign of
the leading ($1/y^5$) term depends on which condition is taken
\cite{dc}.

\subsection{Expectation value of the stress tensor on the brane}
\label{sec:tmnotb}

The maximal symmetry of the $y=const$ surfaces ensures that
$$
\la\partial^i\Psi\partial_j\Psi\ra={1\over
n+1}\,\la\partial^k\Psi\partial_k\Psi\ra\;h^i_j~.
$$
Hence, we can compute the v.e.v. of the surface stress tensor
induced by the bulk field $\Psi$ as
$$
\la T_{i}^{j}\ra^{brane} =-(n-1) r_c\,
\la\partial_i\Psi\partial^j\Psi\ra 
\,\delta(y)
=(n-1) r_c\,   h_{i}^{j} \,\delta(y)\, %
\lim_{\zeta,y,y'\to0}{\partial_\zeta
G^{(1)}(y,y',\zeta)\over2\zeta} .%
$$
As is well known, the stress tensor on the brane diverges even
after subtracting the Minkowski part. So, we have to introduce
some alternative or additional method to regulate it. One option
is to use dimensional regularization. Expressions like $1/y^{n+2}$
are regularized by taking the $y\to0$ limit for negative enough
$n$, and then continuing back to $n=3$.
Using this prescription, we obtain
\beq\label{tonb} %
\la T^i_j \ra^{brane}%
={2(n-1)\over S_{(\q+1)}}
    {\Gamma\lp(-1-n\rp)\over r_c^{n+1} }\;\delta(y)\;h^i_{~j}
={1\over 16 \pi^2}\lp(%
{1\over r_c^4}{1\over n-3}-{\log\lp(r_c\mu\rp)\over r_c^4} \rp)\;
\delta(y)\; h^i_j
\eeq %
where we introduced an arbitrary renormalization scale $\mu$.
The divergence requires a renormalization of $\varphi^4$, where
$\varphi=1/r_c$ is the relevant field for the effective field
theory, as argued in Section \ref{sec:classpot}.

To obtain a more physical picture, we shall show the form of the
brane stress tensor regularized using a cutoff, which we can
assume is related to the brane thickness $d$. To this end, we can
evaluate the point split expression at $y=0$ and $y'=d$, and we
obtain\footnote{Note that once we introduce a cutoff distance, in
order to compute $\tmn$ we should use the full Green function
instead of the subtracted one (\ref{gsub1}).}
\beqn\label{tonbd1} %
\la T^i_j \ra^{brane}%
&=&-{3\over 2\pi^2}
    {e^{d/r_c}\Gamma\lp(-4,d/r_c\rp)\over r_c^{4}
    } \;\delta(y)\;h^i_{~j}\\
\label{tonbd2}&=&-{1\over 16 \pi^2}\lp(%
{6\over d^4}-{2\over d^3r_c}+{1\over
d^2r_c^2}-{1\over dr_c^3}+{\log\lp(r_c\mu\rp)\over r_c^4}
+\dots\rp)\; \delta(y)\; h^i_j 
\eeqn %
where the dots denote positive powers of $d$ and the expansion
(\ref{tonbd2}) holds for $d\ll r_c$. In the opposite limit, Eq.
(\ref{tonbd1}) is perfectly regular, and in fact it vanishes for
$r_c\to0$.
Several comments are now in order. First of all, note that the
finite parts of Eqs. (\ref{tonb}) and (\ref{tonbd2}) agree. Since
Eq. (\ref{tonbd2}) only holds for $r_c\gg d$, we observe that the
dimensional regularization result correctly captures the large
$r_c$ form of $\tmn^{brane}$, and should not be trusted for small
$r_c$.
We shall interpret that the divergent terms in the expansion
(\ref{tonbd2}) should be absorbed in a (finite set of)
counterterms involving negative powers of $r_c$. This goes along
the same lines of Sec. (\ref{sec:boundact}), where we saw that the
field associated to $r_c$ appearing in the 4D effective theory (in
the large-$r_c$ regime) is nothing but $\varphi=1/r_c$.

Before discussing Eqns. (\ref{tonb})--(\ref{tonbd2}) and the
application of these results to the computation of the effective
potential for $r_c$, we shall display the form of the brane stress
tensor when the brane is inflating. Again, dimensional
regularization leads to a result valid for large $r_c$. We shall
not include here the form of $\la T^i_j \ra^{brane}$ using a
cutoff distance. The expressions are lengthy and do not illustrate
essentially anything more than already contained in
(\ref{tonbd2}).
For the $-$ branch, we obtain
\beqn\label{tmnONB-} %
\la T^i_j \ra|_0&=& {(n-1)H^{n+1}\over S_{(\q+1)}}
{2+n Hr_c\over 1+n Hr_c} \;B\lp(n+2+{1\over Hr_c},-n-1\rp)%
\;\delta(y)\;h^i_{~j}\\ %
&=&-{(1+Hr_c)(1+2Hr_c)(2+3Hr_c)(1+4Hr_c)\over
32\pi^2\;r_c^4}\cr%
&&~~~~~~~~~~~~~\times\lp\{{1\over n-3} +\psi\lp(5+{1\over
Hr_c}\rp)+\log\lp(H/\mu\rp) -{Hr_c\over(1+3Hr_c)(2+3Hr_c)}+\dots
\rp\}\;\delta(y)\;h^i_{~j}~,\nonumber
\eeqn %
up to a rescaling of $\mu$.

For the $+$ branch, we obtain
\beqn\label{tmnONB+} %
\la T^i_j \ra|_0&=& {(n-1)H^{n+1}\over S_{(\q+1)}} %
{2-n Hr_c\over 1-n Hr_c} B\lp(2+{1\over Hr_c},-n-1\rp) \;\delta(y) \;h_{i}^{j}\cr %
&=&{(1-Hr_c)(1-2Hr_c)(2-3Hr_c)(1-4Hr_c)\over 32\pi^2\;r_c^4}
\cr %
&&~~~~~~~~~~~~~\times \lp\{{1\over n-3} %
+\psi\lp({1\over Hr_c}-4\rp)+\log\lp(H/\mu\rp) %
+{Hr_c\over(1-3Hr_c)(2-3Hr_c)}+\dots \rp\}\;\delta(y)\;h_{i}^{j}%
\eeqn %

There are five types of divergent terms in (\ref{tmnONB-}) and
(\ref{tmnONB+}), proportional to $H^4$, $H^3 /r_c$, $H^2 /r_c^2$,
$H /r_c^3$, and $1 /r_c^4$. They can be renormalized with
counterterms of the type $R_{(4)}^2$, $K^3 \varphi$, $R_{(4)}
\varphi^2$, $K \varphi^3$ and $\varphi^4$, where $K$ is the
extrinsic curvature and as before $\varphi=1/r_c$. Note as well
that we can readily identify two of the special cases when the
divergent part of (\ref{tmnONB+}) vanishes: one corresponds to the
truly self-accelerated solution (with $\tau=0$, and $Hr_c=1$); the
other corresponds to the only case when Eq. (\ref{junction}) has
only one solution, $Hr_c=1/2$.

In the $H\to0$ limit, the finite part reduces to
$$
\la T^i_j \ra|_0 =\mp{1\over 16\pi^2}\;{\log\lp(\mu r_c\rp)\over
r_c^4}\;h^i_j\;\delta(y)~,
$$
which agrees with the flat brane case for the self-accelerated
branch only. This also happens in the expressions for the
components of the bulk stress tensor Eqns. (\ref{Tyyy=0}) and
(\ref{TyyFlaty=0}). This might seem strange at first sight, but we
shall this might be a consequence that in the $-$ branch the bound
state is present for all nonvanishing values of $H$, so this limit
is discontinuous in this sense.

\section{1-loop effective potential for $r_c$}
\label{sec:veff}

This Section is divided into two parts. In the first, we obtain
the 1-loop effective potential $V_\eff$ for $r_c$ in the flat
brane case, by three methods. The first one uses the expression
that we obtained in Sec \ref{sec:tmnotb} for the (expectation
value of) the stress tensor on the brane. We re-derive it using
the  usual Kaluza-Klein decomposition and, finally, using the
boundary effective theory of Section \ref{sec:boundact}.
Then, we discuss the application of this potential to attempt to
generate a large value of the cross-over scale $r_c$ in Sec
\ref{sec:largerc}.

\subsubsection*{$V_\eff$ obtained from the stress tensor on the brane, Eq. (\ref{tonb})}

From (\ref{action}), the coupling between the (canonical) field
$\varphi$ associated to $r_c$ and $\Psi$ is of the form
$r_c(\varphi) (\partial\Psi)^2$. The equation of motion for
$\varphi$ is
%
%
\beq\label{eomvphi} %
\Box_4 \varphi= r_c'(\varphi)(\partial\Psi)^2~.
\eeq %
To take into account the quantum effects, one can substitute
$(\partial\Psi)^2$ by its vacuum expectation value.
Then, we can identify the 1-loop effective potential
$V_\eff(\varphi_c)$ induced by $\Psi$ as
\beq\label{veffprime} %
V_\eff'(r_c)=\la(\partial\Psi)^2\ra ~,
\eeq %
and using Eq. (\ref{tonbd2}),
\beq\label{veff} %
V_\eff(r_c)={A\Lambda^4}  + B{\Lambda^3\over r_c}+C{\Lambda^2\over
r_c^2} +D{\Lambda\over r_c^3} -{1\over 64\pi^2}
{\log(r_c\Lambda)\over r_c^4}+\dots~,
%
\eeq %
where $A,B,C,D$ are constants, we identified $\Lambda$ as the
cutoff scale which we can associate to the brane thickness,  and
the dots indicate suppressed terms. As argued before, the relevant
field in the 4D effective action is $\varphi=1/r_c$. We discuss
what renormalization conditions and the application of this result
below. Before that, we shall re-derive the same result with two
other methods.

\subsubsection*{Kaluza-Klein decomposition}

We can obtain slightly better understanding of the result
(\ref{veff}) by working in terms of the dimensional or
Kaluza-Klein (KK) reduction. Introducing the KK decomposition
$\Psi=\int dm \, \uu_m \psi_m$ as in Eq. (\ref{kkdecomp}), we see
from Eq. (\ref{veffprime}) that
\beq\label{kk1} %
V_\eff'(r_c)= \int dm \; \uu_m^2 \la (\partial\psi_m)^2 \ra ~,
\eeq %
where $\psi_m$ are 4D scalar fields with mass $m$. In the previous
equation, we have assumed that modes with different $m$ are
independent, so that one has $\la \psi_m \psi_{m'}\ra = \la
\psi_m^2\ra \delta(m-m')$.
We can easily compute $\la (\partial\psi_m)^2 \ra$ {\em e.g.}
using point splitting. One finds
\beq\label{dphi2} %
\la (\partial\psi_m)^2 \ra={1\over 8\pi^2}\lp\{ %
16 \Lambda^4 -4 m^2 \Lambda^2 - m^4 \log\lp(m/\Lambda\rp)+\dots \rp\}~,%
\eeq %
where $\Lambda$ is the cutoff and the dots stand for suppressed
terms.
For a Minkowski brane, the wave function of the KK modes on the
brane is (see Eq. (\ref{uukk}))
$$
\uu_m^2|_0={1\over\pi}{1\over1+(r_c m)^2} ~.
$$
The integral over the KK continuum of (\ref{dphi2}) with this
'weight' contains a few divergent terms. The first term in
(\ref{dphi2}) gives a quartic divergence, while the rest give at
most cubic divergences. As we shall see below this is neatly
captured by the boundary effective theory of Sec
\ref{sec:boundact}. Aside from this, we shall only comment that
this
results in a potential with the same features as (\ref{veff}). In
particular, we reproduce the same coefficient for the finite term.
This agreement might not be surprising since, after all, the Green
function (and so the stress tensor) have also been obtained as a
mode sum. However, the regularization procedure is slightly
different. Eq (\ref{veff}) is obtained by point-splitting the
stress tensor, that is, evaluating the wave-function of the modes
slightly off the brane. Instead, in the method described here, the
sum over the modes is cut-off at a certain frequency.

\subsubsection*{Boundary effective action description}

In the language of the boundary effective action of Sec
\ref{sec:boundact}, the decay of the resonant mode into higher
dimensional modes is described by a (non-local) coupling of the
form $\varphi\,\tilde\psi\Delta\tilde\psi$, where
$\Delta=\sqrt{-\Box_{(4)}}$ and $\varphi=1/r_c$. Then, we can
obtain the 1-loop effective potential as the sum over all 1-loop
1PI diagrams with vanishing the external momentum. One easily
obtains
\beq\label{vefflpr1}%
V_\eff={1\over 2}\int {d^4k\over(2\pi)^4} \log\lp(1+{k/r_c\over
k^2}\rp)~,
\eeq %
where $k^i$ is the Euclidean momentum flowing in the loop. This
integral diverges. Introducing a cutoff $\Lambda$ to regulate it,
we obtain %
\beq\label{vefflpr}%
V_\eff={1\over 64\pi^2}\lp\{%
{4\over3} {\Lambda^3\over r_c}%
-{\Lambda^2\over r_c^2}%
+{4\over3} {\Lambda\over r_c^3}%
-{\log\lp(r_c\Lambda\rp)+1/4 \over r_c^4}%
+\dots \rp\}~,
\eeq%
where the dots denote terms suppressed by powers of $\Lambda$.

\subsection{A large cross-over scale?}
\label{sec:largerc}

We shall regard Eq. (\ref{veff}) as the result for the 1-loop
effective potential $V_\eff$ when the brane is flat (which holds
for large $r_c$). The constants $A,B,$ etc are unknown to the
effective theory, and must be fixed by means of a set of
renormalization conditions. We shall assume that $\varphi=1/r_c$
has a quartic potential $\alpha \varphi^4$ at tree level.
Accordingly, we impose as renormalization conditions that the
coefficients $A, B, C$ and $D$ in (\ref{veff}) vanish. Then, the
potential at 1-loop is of the Coleman-Weinberg \cite{cw} form
\footnote{We could also include the loops of $\varphi$ itself, but
the argument would not change much.}
\beq\label{veffflat} %
V_\eff(\varphi)=\varphi^4\lp(\alpha+\beta\log(\varphi/\Lambda
)\rp)
\eeq %
where $\beta=1/64\pi^2$. The minimum of this potential is at
$r_c\simeq e^{\alpha/\beta}/\Lambda$, which can be very large
without fine tuning. This seems quite encouraging as it might
provide a natural explanation why $r_c$ is so large in the DGP
model, and ultimately why the current acceleration of the universe
is so small. This potential gives a very small mass, of order
$1/r_c$. Aside from being potentially unstable under quantum
corrections, this would require a very suppressed coupling of
$\varphi$ to matter in order not to generate an observable fifth
force \cite{ahn}.
Rather than elaborating on these issues, we shall now address how
this picture changes when we include the curvature of the brane --
an essential ingredient if we are concerned about the accelerated
expansion.

For an inflating brane, the 1-loop contribution (which can be
obtained from (\ref{veffprime}) and (\ref{tmnONB-}),
(\ref{tmnONB+})) is slightly more complicated than
(\ref{veffflat}).
The most important difference is now the classical contribution
$\sim\pm H m_P^2 \,\varphi$ discussed in Section
\ref{sec:classpot}, which clearly dominates over (\ref{veffflat}).
In the following, we shall assume that $\varphi$ is not protected
by any symmetry. It is then 'natural' to have a potential of the
form
$$
V_\eff(\varphi) =\pm H m_P^2 \,\varphi +{A\Lambda^4}  +
B{\Lambda^3 \varphi}+C{\Lambda^2 \varphi^2}+\dots~,
$$
where $A$, $B$, ... are order-one coefficients and $\Lambda$ is
the cutoff.
Given that the piece linear in $\varphi$ represents the term due
to the background curvature we shall take the renormalization
condition $B=0$ (which amounts to a redefinition of $m_P$). Then,
the potential has a minimum (maximum) for the normal
(self-accelerated) branch at $r_c\sim (\Lambda/m_P)^2 \,C/H $.
There are different estimates available in the literature
\cite{dgp,scales,lpr,nr,dvaliir} of the cutoff $\Lambda$ for the
bulk graviton, but all of them are much lower than $m_P$. Hence,
the expectation value generated for $r_c$ is too small for
phenomenological purposes.

As an aside, we shall mention that an intriguing situation arises
if the mass term for $\varphi$ is Planckian. This is the case, for
instance, if this field couples directly to matter and the cutoff
in the matter loops is of order $m_P$. One should still keep the
linear term, as it encodes the coupling to the background
curvature. Then, the extremum occurs at $r_c\sim 1/H$,
\footnote{Note that the 1-loop contribution alone also admits
local minima at $r_c\sim 1/H$.} which seems quite suggestive for
phenomenology. This corresponds to a maximum for the
self-accelerated branch and, in fact, it appears from Eq.
(\ref{junction}) that generically there are no self-accelerated
solutions with $r_c\sim 1/H$. In the normal branch, instead,
$r_c\sim 1/H$ is a minimum. This implies that effectively the
value of $r_c$ 'tracks' the Hubble radius, modifying the whole
setup. An analysis of the cosmology in this kind of scenario is
more complicated (see \cite{wb-l} for the case of a time-dependent
Brans-Dicke field on the brane) and is left for future research.

\section{Conclusions}
\label{sec:conc}

We have discussed the quantum effects from the bulk fields in the
DGP model, in a toy version of it where instead of the graviton we
considered a bulk scalar field with a kinetic term on the brane.
We have computed the Green function and the stress tensor induced
by quantum fluctuations. We have shown that they are regular
everywhere except on the brane, and described their form in
several limits of interest.

The cross-over scale $r_c$ in the DGP model is usually assigned a
large value --of order of the present Hubble radius -- in a
somewhat {\em ad hoc} way. We used our results to clarify whether
the quantum effects can provide a natural mechanism to generate
such a large value. Once $r_c$ is promoted to a dynamical field,
this problem translates to finding the potential that drives it to
large values. Here, we considered the 1-loop effective potential
$V_\eff(r_c)$ induced by the bulk fields.

With a flat brane in a flat bulk, we showed that this is possible
by a quite simple argument. The fluctuations of the five
dimensional 'graviton' provide a Coleman-Weinberg-type potential,
which can induce a minimum at very large $r_c$ without fine
tuning.
When the curvature of the brane is included, the potential is
largely modified due to a tree-level contribution originating from
the background curvature.
In the present setup, the value thus generated for $r_c$ is too
small to be compatible with observations. However, simple
extensions of the model suggest that it is possible to generate a
minimum where $r_c$ tracks the Hubble radius $1/H$ in the normal
branch.

\acknowledgements

We thank Takahiro Tanaka for many useful comments, Gia Dvali,
Gregory Gabadadze, Jaume Garriga, Jose Juan Blanco-Pillado,
Michele Redi, Alberto Iglesias, Eduardo Pont{\' o}n for helpful
conversations and the Center for Cosmology and Particle Physics
for kind hospitality during the realization of this article. We
acknowledge support from the Departament d'Universitats, Recerca i
Societat de la Informaci{\'o} of the Generalitat de Catalunya,
under the Beatriu de Pin{\' o}s Fellowship 2005 BP-A 10131.

\appendix

\section{Some formulas for the Green function}
\label{app:green}

In order to perform the integral in (\ref{gsub}), it is convenient
to use the representation for the massive Hadamard function in dS
space \cite{pt}
\begin{eqnarray*}
 G^{(dS)(1)}_p(x,x')&=&{2\over (n-1)S_{(n)}}%
 {2^{-{n-1\over2}}\over(1-\cos\zeta)^{\q-1\over 2}}
      \Biggl\{ {}_2F_1\left(-ip+{1\over 2},ip+{1\over 2},{-\q+3\over 2};
     {1-\cos\zeta\over 2}\right)\cr %
     &+&
     {\Gamma\left({\q\over 2}-ip\right)\Gamma\left({\q\over 2}+ip\right)
          \Gamma\left(-{\q-1\over 2}\right)
     \over \Gamma\left({1\over 2}-ip\right)\Gamma\left({1\over 2}+ip\right)
          \Gamma\left({\q-1\over 2}\right)
      }\left({1-\cos\zeta\over 2}\right)^{\q-1\over 2}
      {}_2F_1\left(-ip+{\q\over 2},ip+{\q\over 2},{\q+1\over 2};
     {1-\cos\zeta\over 2}\right)
   \Biggr\}~,
\end{eqnarray*}
where ${}_2F_1$ is the hypergeometric function and $\zeta$ is the
invariant distance between $x$ and $x'$.

Including the bound state contribution, we obtain
\begin{align}
\label{bulk}
  \overline{G}^{(1)}
   ={ 1\over   S_{(\q)}\Gamma\left({n+1\over
   2}\right)} {\lp(1\over 4a a' \rp)^{n/2}}%
    \; \sum_{\s=0}^{\infty}    \sum_{j=0}^{\infty} %
&\lp\{-1-2 {\lambda_>+\lambda_<\over \lambda_>-\lambda_<}\lp(
{\lambda_>\over n/2+k+j-\lambda_>} -{\lambda_<\over
n/2+k+j-\lambda_<}\rp)\rp\}\times\cr
          &{(-1)^\s \Gamma\left(\q+2\s+j\right)
            \over j!\,\s !\, \Gamma\left({n+1\over 2}+\s\right)}
          \lp(aa'H^2 \rp)^{\mp(n/2+\s+j) }
          \left({1-\cos\zeta\over 2}\right)^\s~.%
\end{align}
where $\zeta$ is the invariant distance in $dS$ space and
$S_{(n+1)}=2 \pi^{1+n/2}/\Gamma(1+n/2)$ is the volume of a unit
$n+1$ dimensional sphere. This sum can be readily performed for
the first term in the curly brackets, leading to minus $G_0$ as
defined in (\ref{G0}).
%
%
%
The two remaining terms in (\ref{bulk}) can be obtained from $G_0$
as
\beq \label{op} %
G_{\lambda}= \int_0^\infty d \alpha \, e^{\lambda \alpha } \; %
G_0\lp(H\to e^{\pm\alpha/2} H\rp) ~,%
\eeq %
which leads to (\ref{op2}) quite immediately.


~\\

For the sake of completeness, we include here the expression of
the Green function when $x$ and $x'$ are 'aligned', that is, when
their four dimensional coordinates coincide, $\zeta=0$. We obtain
\beqn \label{Gresult}
G^{(1)}= -{H^n \,(aa' H^2)^{\pm n/2}\over n S_{(n+1)}} %
\Biggl\{      {(aa' H^2)^{n}\over |(aa' H^2)^2-1|^n}
+2{\lambda_>+\lambda_<\over \lambda_>-\lambda_<}\lp[%
A\lp(\lambda_>\, ,  (aa' H^2)^{\mp1} \rp) %
%
- %
A\lp(\lambda_<\, ,  (aa' H^2)^{\mp1} \rp)
 \rp] \Biggr\}~,
\eeqn where we introduced
$$
A\lp(\lambda,x\rp)\equiv
\lambda \; x^{\lambda}\; B_{x}\lp( {n\over2}-\lambda,1-n  \rp) %
$$
and $B_x(a,b)$ is the incomplete Beta function. For small $x$,
\beq\label{Hbehaviour} %
A\lp(\lambda,x\rp)\simeq  {\lambda\over n/2-\lambda}\; x^{n/2}
\lp(1+{\cal O}(x)\rp)~.
\eeq %

Let's now elaborate on the IR divergence present in
(\ref{Gresult}) for the $-$ branch, due to the $\lambda_>=n/2$
contribution (the massless bound state).
Using
$$
B_{x}\lp(a,b\rp) = (-1)^{-a} %
B\lp(1-a-b,a \rp)
+ (-1)^{b} B_{1/x}\lp( 1-a-b,b \rp)
$$
where $B(a,b)=B_1(a,b)$ is the Euler Beta function,  we can find
that the expansion around $\lambda= n/2$,
$$
{e^{n Z /2}\over\lambda}\; A\lp(\lambda,x\rp)\simeq   -{1\over \lambda  -n/2}%
-\log x  + (-1)^{n+1} B_{x}\lp(n, 1-n\rp) - \psi\lp(n\rp) -\gamma
+ i \pi + {\cal O}\lp(\lambda-n/2\rp)~,
$$
up to an irrelevant constant, and $\psi(z)=\Gamma'(z)/\Gamma(z)$
is the digamma function. In this equation, we kept the digmma term
in order to obtain a finite limit for $n=3$. For small $x$,
$$
(-1)^{n+1} B_{1/x}\lp( n,1-n\rp) - \psi\lp(n\rp) \sim  %
\gamma -i\pi+\log x + n x +\dots 
$$
up to constant terms and where the dots denote higher powers of
$x$. The terms linear in $\log x$ cancel, and the leading order
term is ${\cal O}\lp(x\rp)$. The other two contributions to the
Green function behave like $const + {\cal O}\lp(x\rp)$. Hence, the
Green function is well behaved at infinity (which is represented
by the light cone in the $-$ branch).


\end{document}